# NEW MULTIFERROICS BASED ON $Eu_xSr_{1-x}TiO_3$ NANOTUBES AND NANOWIRES

Eugene A. Eliseev[1], Maya D. Glinchuk[1], Victoria V. Khist[1], Chan-Woo Lee[2], Chaitanya S. Deo[3], Rakesh K. Behera[3*], and Anna N. Morozovska [4†]

[1] Institute for Problems of Materials Science, NAS of Ukraine,
Krjijanovskogo 3, 03142 Kiev, Ukraine

[2] The Makineni Theoretical Laboratories, Department of Chemistry, University of Pennsylvania,
Philadelphia, PA 19104, USA

[3] Nuclear and Radiological Engineering Program, George W. Woodruff School of Mechanical Engineering, Georgia Institute of Technology,
Atlanta, GA 30332, USA

[4] Institute of Physics, NAS of Ukraine, 46, pr. Nauki, 03028 Kiev, Ukraine

Using Landau-Ginzburg-Devonshire theory we have addressed the complex interplay between structural antiferrodistortive order parameter (oxygen octahedron rotations), polarization and magnetization in $Eu_xSr_{1-x}TiO_3$ nanosystems. We have calculated the phase diagrams of $Eu_xSr_{1-x}TiO_3$ bulk, nanotubes and nanowires, which include the antiferrodistortive, ferroelectric, ferromagnetic and antiferromagnetic phases. For $Eu_xSr_{1-x}TiO_3$ nanosystems, our calculations show the presence of antiferrodistortive-ferroelectric-ferromagnetic phase or the triple phase at low temperatures ($\leq 10$ K). The polarization and magnetization values in the triple phase are calculated to be relatively high ($\sim 50$ $\mu C/cm^2$ and $\sim 0.5$ MA/m). Therefore, the strong coupling between structural distortions, polarization and magnetization suggest the $Eu_xSr_{1-x}TiO_3$ nanosystems as strong candidates for possible miltiferroic applications.

---

[*] Corresponding author 1: rkbehera@ufl.edu

[†] Corresponding author 2: morozo@i.com.ua

# I. INTRODUCTION

The search for new multiferroic materials with large magnetoelectric (ME) coupling are very interesting for fundamental studies and important for applications based on the magnetic field control of the material dielectric permittivity, information recording by electric field, and non-destructive readout by magnetic field [1, 2]. Solid solutions of different quantum paraelectrics (such as $Eu_xSr_{1-x}TiO_3$ or $Eu_xCa_{1-x}TiO_3$) subjected to elastic strains can be promising for multiferroic applications.

Bulk $SrTiO_3$ is nonmagnetic quantum paraelectric at all temperatures [3]. Below 105 K bulk $SrTiO_3$ has antiferrodistortive (AFD) structural order [4, 5, 6], characterized by spontaneous oxygen octahedron rotation angles (or "tilts"), which can be described by an axial vector $\Phi_i$ ($i$=1, 2, 3) [7]. However, $SrTiO_3$ thin films under misfit strain are ferroelectric up to 400 K [8].

Bulk quantum paraelectric $EuTiO_3$ is AFD below about 281 K [9, 10, 11, 12], antiferromagnetic (AFM) at temperatures lower than 5.5 K and paraelectric at all other temperatures [1, 2]. Using first principles calculations, Fennie and Rabe [13] predicted the presence of simultaneous ferromagnetic (FM) and ferroelectric (FE) phases in (001) $EuTiO_3$ thin films under compressive epitaxial strains exceeding 1.2%. They demonstrated that strain relaxation to the values lower than 1% eliminates FE-FM phase appearance in $EuTiO_3$ thin film [13]. Lee *et al.* [14] demonstrated experimentally that $EuTiO_3$ thin films with thickness 22 nm on $DyScO_3$ substrate become FM at temperatures lower than 4.24 K and FE at temperatures lower than 250 K under the application of more than 1% tensile misfit strain.

The intrinsic surface stress can induce ferroelectricity, ferromagnetism and increase corresponding phase transition temperatures in conventional ferroelectrics and quantum paraelectric nanorods, nanowires [15, 16, 17, 18, 19] and binary oxides [20]. The surface stress is inversely proportional to the surface curvature radius and directly proportional to the surface stress tensor (similar to Laplace surface tension). Thus, the intrinsic surface stress should depend both on the growth conditions and the surface termination morphology [21, 22]. Surface reconstruction should affect the surface tension value or even be responsible for the appearance of surface stresses [23, 24]. Using Landau-Ginzburg-Devonshire (LGD) theory Morozovska et al. [25] predicted the FE-FM multiferroic properties of $EuTiO_3$ nanowires originated from the intrinsic surface stress. However, the important impact of the structural AFD order parameter (oxygen octahedron rotations) in $EuTiO_3$ has not been considered so far. Since the AFD order parameter strongly influences the phase diagrams, polar and pyroelectric properties of quantum paraelectric $SrTiO_3$ [26, 27, 28, 29], similar influence is expected for $EuTiO_3$ and $Eu_xSr_{1-x}TiO_3$. Therefore, a fundamental study of the possible appearance of the polar, magnetic and multiferroic phases in AFD $Eu_xSr_{1-x}TiO_3$ solid solution system seems necessary. Recently the transition from paraelectrics cubic phase to AFD phase in solid solution $Eu_xSr_{1-x}TiO_3$ has been studied by means of Electron Paramagnetic Resonance [30].

In this work we study the possibility of inducing simultaneous ferroelectricity and ferromagnetism in $Eu_xSr_{1-x}TiO_3$ nanosystems within conventional LGD theory allowing structural ordering. **Figure 1** illustrates the nanosystems considered in this study, (a) nanotubes clamped to the rigid core, where the outer sidewall of the tube is mechanically free and electrically open, i.e., non-electroded (**Fig. 1a**) and freestanding nanowires (**Fig. 1b**) In the nanotube cases, technologically convenient materials for a rigid core can be ZnO, Si, SiC ultra-thin nanowires. Perovskite-type cores like $LaAlO_3$, $LaAlSrO_3$, $DyScO_3$, $KTaO_3$ or $NdScO_3$ are more sophisticated to design. Since the lattice constants of $EuTiO_3$, $SrTiO_3$, and $Eu_xSr_{1-x}TiO_3$ in the cubic phase are similar (~ 3.905 Å), the misfit strains due to the rigid core for $Eu_xSr_{1-x}TiO_3$ nanotubes will be similar to $EuTiO_3$ nanotubes. Therefore, the misfit strains appeared at the $Eu_xSr_{1-x}TiO_3$ tube-core interface are approximately -6% for ZnO, -1.7% for Si, 10% for SiC, -4% for $LaAlSrO_3$, -3% for $LaAlO_3$, +0.9% for $DyScO_3$, +2.1% for $KTaO_3$ and +2.6% for $NdScO_3$ core. In this study we considered the axial polarization $P_3$ directed along the tube axis $z$, while the radial polarization $P_\rho$ perpendicular to the surface of the tube is neglected due to the strong depolarization field $E_\rho^d \sim -P_\rho/\varepsilon_0\varepsilon_b$ that appears for the component in the case of non-electroded tube sidewalls [25].

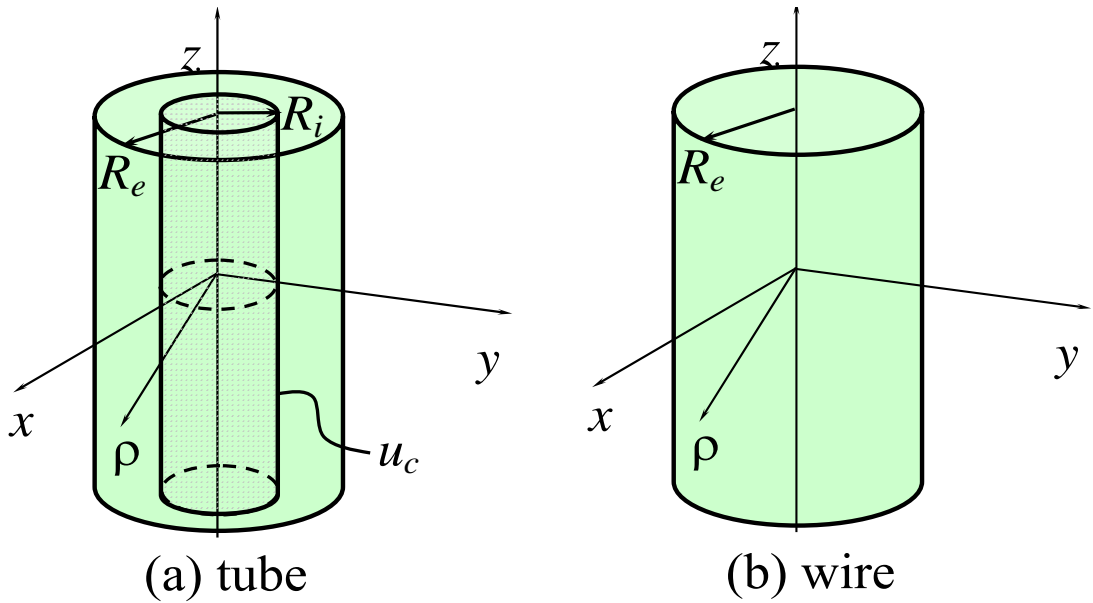

**Figure 1.** (a) Schematics of a nanotube clamped on a rigid core. Mismatch strain $u_c$ can exist at the tube-core interface. The tube outer radius is $R_e$, the inner radius is $R_i$, $\rho$ is the polar radius. (b) Schematics of a nanowire.

The application of the continuum media LGD theory and intrinsic surface stress conception to the description of nanosized particles polar and magnetic properties requires justification due to the small size of the object of study. The continuum media theory was successfully used for the analysis of elastic properties of metallic, semiconductor, dielectric or polymeric nanowires and nanotubes [31, 32, 33, 34, 35, 36, 37] and the piezoelectric response [38]. For nanosized ferroics, the applicability of the continuous media phenomenological theory is corroborated by the fact that the critical sizes (~2-10 lattice constants) of the appearance of long-range order calculated from atomistic [39, 40] and phenomenological theories [17, 18, 19] are in good agreement with each other [37, 41, 42, 43, 44] as well as with experimental results [45]. Generally, the long-range order appears for sizes larger than the critical sizes (details can be found in Refs. [17-20, 25]). Once the long-range order is established, it is possible to apply the mean field LGD theory [46]. Thus the agreement between the magnitudes of the critical sizes calculated from LGD and atomistic theories are extremely important.

This paper is organized as follows: the basic equations of LGD-theory and material parameters of $Eu_xSr_{1-x}TiO_3$ used in the calculations are listed in Section II. The predicted phase diagrams of $Eu_xSr_{1-x}TiO_3$ bulk, nanowires and nanotubes are presented and analyzed in Section III. The variation of spontaneous polarization and magnetization in $Eu_xSr_{1-x}TiO_3$ nanosystems are presented in Section IV. The results are discussed in Section V and the details of the fitting procedure to obtain the parameters for the LGD calculations are given in the Supplementary Materials.

## II. LANDAU-GINZBURG-DEVONSHIRE THEORY FOR $Eu_xSr_{1-x}TiO_3$

The LGD free energy density $G$ of $Eu_xSr_{1-x}TiO_3$ solid solution depends on the polarization vector **P**, oxygen octahedra tilt vector **Φ**, magnetization vector **M** and antimagnetization vector **L** as:

$$G = G_{grad} + G_S + G_{elastic} + G_{ME} + G_M + G_{P\Phi} \quad (1)$$

where $G_{grad}$ is the gradient energy, $G_S$ is the surface energy, $G_{elastic}$ is elastic energy, $G_{ME}$ is the biquadratic ME energy, $G_M$ is magnetization-dependent energy, and $G_{P\Phi}$ is polarization-and-tilt-dependent energy.

The form of $G_{grad} + G_S$ is the same as listed in Ref.[25]. The elastic energy is given as $G_{elastic} = -s_{ijkl}\sigma_{ij}\sigma_{kl}/2$, where elastic compliances $s_{ijkl}(x) = xs_{ijkl}^{EuTiO_3} + (1-x)s_{ijkl}^{SrTiO_3}$; $\sigma_{ij}$ is the elastic stress tensor. The biquadratic ME coupling energy density ($G_{ME}$) is given as

$$G_{ME} = \int_V d^3r \frac{P_3^2}{2}\left(\gamma_{FM}M^2 + \gamma_{AFM}L^2\right) \quad (2)$$

Here $P_3$ is FE polarization, $M^2 = M_1^2 + M_2^2 + M_3^2$ is the square of FM magnetization, and $L^2 = L_1^2 + L_2^2 + L_3^2$ is the square of AFM order parameter vector.

Magnetic properties are observed in $EuTiO_3$ and are absent in $SrTiO_3$. Therefore, composition dependence of the biquadratic ME coupling coefficients $\gamma_{FM}(x)$ and $\gamma_{AFM}(x)$ should be included into Eq.(2). Here we assume a linear dependence on Eu content ($x$) above percolation threshold ($x_{cr}$) (see e.g. Ref. [47]), namely $\gamma_{AFM}(x) = \gamma_{AFM}^{EuTiO_3}(x - x_{cr}^A)/(1 - x_{cr}^A)$ and $\gamma_{FM}(x) = \gamma_{AFM}^{SrTiO_3}(x - x_{cr}^F)/(1 - x_{cr}^F)$ at content $x_{cr}^{F,A} \leq x \leq 1$; while $\gamma_{AFM}(x) = 0$ and $\gamma_{FM}(x) = 0$ at $x < x_{cr}^{A,F}$. The percolation threshold concentration $x_{cr}$ can be estimated from the percolation theory [47]. For the simple cubic sub-lattice of magnetic ions (Eu) $x_{cr}^F \approx 0.24$ [47], while the percolation threshold is supposed to be higher for AFM ordering, $x_{cr}^A \approx 0.48$ (see e.g. [48]). Note that superscripts "F" and "A" in $x_{cr}^{F,A}$ designate the critical concentrations related to FM and AFM ordering respectively. Following Lee *et al.* [14] we can regard that $\eta_{AFM}^{EuTiO_3} \approx -\eta_{FM}^{EuTiO_3} > 0$ for numerical calculations, as anticipated for equivalent magnetic Eu ions with antiparallel spin ordering in a bulk EuTiO$_3$.

The magnetization-dependent part of the free energy is [20, 25]:

$$G_M = \int_V d^3r \left( \frac{\alpha_M}{2} M^2 + \frac{\alpha_L}{2} L^2 + \frac{\beta_M}{4} M^4 + \frac{\beta_L}{4} L^4 + \frac{\lambda}{2} L^2 M^2 - \sigma_{mn} \left( Z_{mnkl} M_k M_l + \tilde{Z}_{mnkl} L_k L_l \right) \right) \quad (3)$$

where, coefficient $\alpha_M(T,x) = \alpha_C(T - T_C(x))$, $T$ is absolute temperature, $T_C(x) = T_C^0(x - x_{cr}^F)/(1 - x_{cr}^F)$ is the solid solution FM Curie temperature defined at $x_{cr}^F \leq x \leq 1$. $T_C^0$ is the Curie temperature for bulk EuTiO$_3$. Note, that $\alpha_M(T, x = 0) = \alpha_C(T - T_C)$ determines the experimentally observed inverse magnetic susceptibility in paramagnetic phase of EuTiO$_3$ [1, 2, 14]. Also, coefficient $\alpha_L(T,x) = \alpha_N(T - T_N(x))$, where Néel temperature $T_N(x) = T_N^0(x - x_{cr}^A)/(1 - x_{cr}^A)$ is defined at $x_{cr}^A \leq x \leq 1$. $T_N^0$ is the Néel temperature for bulk EuTiO$_3$. The magnetic Curie and Néel temperatures are zero at $x < x_{cr}^{F,A}$. For equivalent amount of magnetic Eu ions with antiparallel spin ordering it can be assumed that $\alpha_C \sim \alpha_N$. The positive coupling term $\frac{\lambda}{2} L^2 M^2$ prevents the appearance of FM as well as ferrimagnetic (FiM) phases at low temperatures ($T < T_C$) under the condition $\sqrt{\beta_M \beta_L} < \lambda$. Coefficients $\beta_M$, $\beta_L$, $\lambda$ are regarded $x$-independent. $Z_{mnkl}$ and $\tilde{Z}_{mnkl}$ represents magnetostriction and antimagnetostriction tensors respectively.

The polarization and structural parts of the free energy bulk density for cubic m3m parent phase is

$$G_{P\Phi} = \int_V d^3r \left( \frac{\alpha_P}{2} P_3^2 + \frac{\beta_P}{4} P_3^4 - Q_{ij33} \sigma_{ij} P_3^2 + \frac{\alpha_\Phi}{2} \Phi_3^2 + \frac{\beta_\Phi}{4} \Phi_3^4 - R_{ijkl} \sigma_{ij} \Phi_k \Phi_l + \frac{\eta_{i3}}{2} \Phi_i^2 P_3^2 \right) \quad (4)$$

Here $P_i$ is the polarization vector, and $\Phi_i$ is the structural order parameter (rotation angle of oxygen octahedron measured as displacement of oxygen ion). The biquadratic coupling between the structural order parameter $\Phi_i$ and polarization components $P_i$ are defined by the tensor $\eta_{ij}$. [26, 49]. The biquadratic coupling term was later regarded as Houchmandazeh-Laizerowicz-Salje (HLS) coupling [50]. The coupling was considered as the reason of magnetization appearance inside the ferromagnetic domain wall in non-ferromagnetic media [51]. Biquadratic coupling tensor and higher order expansion coefficients are regarded composition dependent: $\beta_{P,\Phi}(x) = x\beta_{P,\Phi}^{EuTiO_3} + (1-x)\beta_{P,\Phi}^{SrTiO_3}$, $\eta_{ij}(x) = x\eta_{ij}^{EuTiO_3} + (1-x)\eta_{ij}^{SrTiO_3}$. $Q_{ijkl}(x) = xQ_{ijkl}^{EuTiO_3} + (1-x)Q_{ijkl}^{SrTiO_3}$ and $R_{ijkl}(x) = xR_{ijkl}^{EuTiO_3} + (1-x)R_{ijkl}^{SrTiO_3}$ are the electrostriction and rotostriction tensors components respectively, which also depend linearly on the composition $x$. Coefficients $\alpha_P(T,x)$ and $\alpha_\Phi(T,x)$ depend on temperature in accordance with Barrett law [52] and composition $x$ of $Eu_xSr_{1-x}TiO_3$ solid solution as $\alpha_P(T,x) = x\alpha_P^{EuTiO_3}(T) + (1-x)\alpha_P^{SrTiO_3}(T)$ and $\alpha_\Phi(T,x) = x\alpha_\Phi^{EuTiO_3}(T) + (1-x)\alpha_\Phi^{SrTiO_3}(T)$ and $\alpha_m(T) = (T_q^m/2)(\coth(T_q^m/2T) - \coth(T_q^m/2T_c^m))$, where sub- and superscript $m = P, \Phi$. Temperatures $T_q^m$ are so called quantum vibration temperatures for SrTiO$_3$ and EuTiO$_3$ respectively, related with either polar (P) or oxygen octahedron rotations ($\Phi$) modes, $T_c^m$ are the "effective" Curie temperatures corresponding to polar soft modes in bulk EuTiO$_3$ and SrTiO$_3$. Note, that recently Zurab Guguchia et al. [30] experimentally observed a nonlinear composition dependence of temperature of transition from cubic non-AFD and tetragonal AFD phase. We have neglected nonlinearity in transition temperatures in this study, since the deviations from the linear dependence is not substantial.

For tetragonal ferroelectric, (anti)ferromagnetic and cubic elastic symmetry groups, coefficients α are renormalized by the surface tension [15-17], misfit strains [53] and biquadratic coupling with a structural order parameter [27-29]. For considered geometry, the renormalization is:

$$\alpha_{RP}(T,x) = \alpha_P(T,x) + \frac{4Q_{12}(x)\mu}{R_e} - \frac{Q_{11}(x) + Q_{12}(x)}{s_{11}(x) + s_{12}(x)} \frac{R_i^2}{R_e^2} u_c - \eta_{11}(x)\frac{\alpha_\Phi(T,x)}{\beta_\Phi}, \quad (5a)$$

$$\alpha_{MR}(T,R,x) \approx \alpha_C\left(T - \left(T_C^0 - \frac{W}{\alpha_C}\frac{4\mu}{R_e} + \frac{Z_{11} + Z_{12}}{\alpha_C(s_{11}(x) + s_{12}(x))} \frac{R_i^2}{R_e^2} u_c\right)\frac{x - x_{cr}^F}{1 - x_{cr}^F}\theta(x - x_{cr}^F)\right), \quad (5b)$$

$$\alpha_{LR}(T,R,x) \approx \alpha_N\left(T - \left(T_N^0 - \frac{\widetilde{W}}{\alpha_C}\frac{4\mu}{R_e} + \frac{\widetilde{Z}_{11} + \widetilde{Z}_{12}}{\alpha_C(s_{11}(x) + s_{12}(x))} \frac{R_i^2}{R_e^2} u_c\right)\frac{x - x_{cr}^A}{1 - x_{cr}^A}\theta(x - x_{cr}^A)\right). \quad (5c)$$

In Eqs. 5, $R_e$ is the tube outer radius, $R_i$ is the tube inner radius (see **Fig. 1**); $\mu$ is the surface tension coefficient, that is regarded as positive; and $u_c$ is misfit strain at the tube-core interface. For the practically important case of the ferroelectric tube deposited on a rigid dielectric core, the tube and core lattices mismatch or the difference of their thermal expansion coefficients determines $u_c$ value allowing for the possible strain relaxation for thick tubes.

If the spontaneous (anti)magnetization is directed along z-axes, the parameters in Eqs.(5b-c) are $\widetilde{W} = +\widetilde{Z}_{12}$, $W = +Z_{12}$, where $Z_{ij}$ and $\widetilde{Z}_{ij}$ are the magnetostriction and anti-magnetostriction coefficients. When the spontaneous (anti)magnetization is along the {x,y} plane, the parameters are $\widetilde{W} = -(\widetilde{Z}_{12} + \widetilde{Z}_{11})/2$, $W = -(Z_{12} + Z_{11})/2$ [25]. Function $\theta(x - x_{cr})$ is the Heaviside step-function [54], i.e. $\theta(x \geq 0) = 1$ and $\theta(x < 0) = 0$. Notice that it is possible to consider radial magnetization, since the influence of demagnetization field existing for such case is typically negligibly small [46].

The terms in Eq.(5) proportional to $\mu/R_e$ originated from the intrinsic surface stress, while the terms proportional to $u_c R_i^2 / R_e^2$ are the strains induced by the rigid core. The size, misfit strain and composition dependence of the ordered phases stability can be obtained from the condition $\alpha_R(T,x) < 0$. In particular, the term $4Q_{12}(x)\mu/R_e$ in Eq.(5a) is negative because $Q_{12}(x) < 0$; so it leads to a reduction in $\alpha_{RP}(T,x)$ and thus favors FE phase appearance for small $R_e$. Since $Q_{11}(x) + Q_{12}(x) > 0$, the term $\sim (Q_{11}(x) + Q_{12}(x))(R_i^2/R_e^2)u_c$ in Eq.(5a) leads to a reduction in $\alpha_{RP}(T,x)$ and thus favors FE phase appearance for positive $u_c$. Similarly, positive terms $-4Z_{12}\mu/R_e$ and $\sim (Z_{11} + Z_{12})(R_i^2/R_e^2)u_c$ favors the appearance of FM phase.

The numerical values of material parameters (Eqs. 1-5) used in the LGD model are listed in **Tables 1** and **2**. The fitting procedures used to estimate the EuTiO$_3$ material parameters are mentioned in the **Supplemental Material**.

Table 1. List of parameters for **polarization and tilt dependent part of the free energy**

| Parameter (SI units) | SrTiO$_3$ Value | Source and notes | EuTiO$_3$ Value | Source and notes |
|---|---|---|---|---|
| $\varepsilon_b$ | 43 | Ref. [55, 56] | 33 | fitting to [1, 57] |
| $\alpha_T^{(P)}$ ($\times 10^6$ m/(F K)) | 0.75 | [4, 26] | 1.95 | fitting to [1, 57] |
| $T_c^{(P)}$ (K) | 30 | [4, 26] | -133.5 | fitting to [1, 57] |
| $T_q^{(P)}$ (K) | 54 | [4, 26] | 230 | fitting to [1, 57] |
| $a_{11}^\sigma$ ($\times 10^9$ m$^5$/(C$^2$F)) | 1.724 | [4, 26] | 1.6 | fitting to [13] |
| $Q_{ij}$ (m$^4$/C$^2$) | $Q_{11}$=0.046, $Q_{12}$=−0.014 | Recalculated from [4] | $Q_{11}$=0.10, $Q_{12}$=−0.015 | fitting to [14] |
| $\alpha_T^{(\Phi)}$ ($\times 10^{26}$ J/(m$^5$ K)) | 18.2 | [4] | 3.91 | fitting to [9, 10] |
| $T_c^{(\Phi)}$ (K) | 105 | [4] | 270 | Averaged value of exp. 220 [10], 275 [30] and 282 [9] |
| $T_q^{(\Phi)}$ (K) | 145 | [4] | 205 | fitting to [10] |
| $\beta_\Phi$ ($\times 10^{50}$ J/m$^7$) | 6.76 | [4] | 1.74 | fitting to [9, 10] |
| $R_{ij}$ ($\times 10^{18}$ m$^{-2}$) | $R_{11}$=8.82, $R_{12}$=-7.77 | recalculated from [4] | $R_{11}$=5.46, $R_{12}$=− 2.35 | fitting to [10] |
| $\eta_{11}$ ($\times 10^{29}$ (F m)$^{-1}$) | 4.19 | [4] | −4.45 | fitting to [1, 57] |
| $s_{ij}$ ($\times 10^{-12}$ m$^3$/J) | $s_{11}$=3.52, $s_{12}$=−0.85 | recalculated from [4, 26] | $s_{11}$=3.65, $s_{12}$=−0.85 | first-principles [58] |

**Table 2. List of parameters for magnetic part of the free energy for EuTiO$_3$**

| LGD-coefficient $\alpha_C \approx \alpha_N$ | Henri/(m·K) | $2\pi \cdot 10^{-6}$ | [1, 14] EXP |
|---|---|---|---|
| LGD-coefficient $\beta_M$ | J m/A$^4$ | $0.8 \times 10^{-16}$ | fitting results of [1] |
| LGD-coefficient $\beta_L$ | J m/A$^4$ | $1.33 \times 10^{-16}$ | fitting results of [1] |
| LGD-coefficient $\lambda$ | J m/A$^4$ | $1.0 \times 10^{-16}$ | fitting results of [1] |
| Magnetostriction coefficients $Z_{ij}$ (Voigt notation) | m$^2$/A$^2$ | $Z_{12} = -(7.5 \pm 0.3) \times 10^{-16}$<br>$Z_{11} = (11.9 \pm 0.3) \times 10^{-16}$ | fitting results of [14] |
| Magnetostriction coefficients $\widetilde{Z}_{ij}$ (Voigt notation) | m$^2$/A$^2$ | $\widetilde{Z}_{12} = -(8.7 \pm 0.2) \times 10^{-16}$<br>$\widetilde{Z}_{11} = (9.2 \pm 0.2) \times 10^{-16}$ | fitting results of [14] |
| AFM Neel temperature $T_N$ | K | 5.5 | [1] |
| FM Curie temperature $T_C$ | K | $3.5 \pm 0.3$ | [1, 2] |
| Biquadratic ME coupling coefficient $\gamma_{AFM} = -\gamma_{FM}$ | J m$^3$/(C$^2$ A$^2$) | $0.08 \times 10^{-3}$ | fitting results of [1] |

To make size effects pronounced, nanosystem sizes should vary from several lattice constants to several tens of lattice constants (*lc*). Our previous analysis showed that size effects can be neglected for nanosystems with sizes of more than 100 *lc* [15-17, 20, 25]. To illustrate typical results for our numerical simulations we have shown mainly the results of three different Eu$_x$Sr$_{1-x}$TiO$_3$ nanosystems: (i) nanotube with small inner radius $R_i = 10$ *lc*, thickness $d = 5$ *lc*, which results in an outer radius $R_e = R_i + d = 15$ *lc*; (ii) nanotube with high inner radius $R_i = 100$ *lc*, thickness $d = 5$ *lc* and outer radius of 115 *lc* (this represents thin film with in-plane polarization); and (iii) a special case of nanotube with $R_i = 0$ *lc*, thickness $d = 5$ *lc*, i.e. a nanowire with radius $R_e = 5$ *lc*. In addition we have also calculated the bulk phase diagram of Eu$_x$Sr$_{1-x}$TiO$_3$ solid solution for comparison. Following Ref.[14], modern epitaxial methods allow to vary misfit strain $u_c$ in the range − 5% to + 5%. Therefore, we will consider the misfit strains within this range for our calculation. Due to the lack of experimental measurements of the surface tension coefficients for Eu$_x$Sr$_{1-x}$TiO$_3$, we have considered μ = 30 N/m based on experimental data for ferroelectric ABO$_3$ perovskites (36.6 N/m [59] or even ~50 N/m [60] for PbTiO$_3$, 2.6-10 N/m for PbTiO$_3$ and BaTiO$_3$ nanowires [61], 9.4 N/m for Pb(Zr,Ti)O$_3$ [62]).

### III. PHASE DIAGRAMS OF Eu$_x$Sr$_{1-x}$TiO$_3$ BULK AND NANOSYSTEMS

Before discussing the phase diagrams for the Eu$_x$Sr$_{1-x}$TiO$_3$ nanosystems, it is necessary to establish the phase diagram of bulk Eu$_x$Sr$_{1-x}$TiO$_3$ system for comparison. **Figure 2a** shows the predicted phases for Eu$_x$Sr$_{1-x}$TiO$_3$ bulk solid solution. The phase diagram shows the presence of 5 different phases: Para (paraelectric-paramagnetic), AFD (antiferrodistortive), AFD-FM (antiferrodistortive-ferromagnetic), AFD-AFM (antiferrodistortive-antiferromagnetic), and AFD-FiM (antiferrodistortive-ferrimagnetic). The magnetic phases AFD-FM, AFD-AFM and AFD-FiM exist at temperatures lower than 10 K. Note that there was no directional control of the polarization and magnetization direction for the bulk system, e.g. our calculations for the bulk solid solution do not differentiate between in-plane, out-of-plane or mixed- ferroelectric phases.

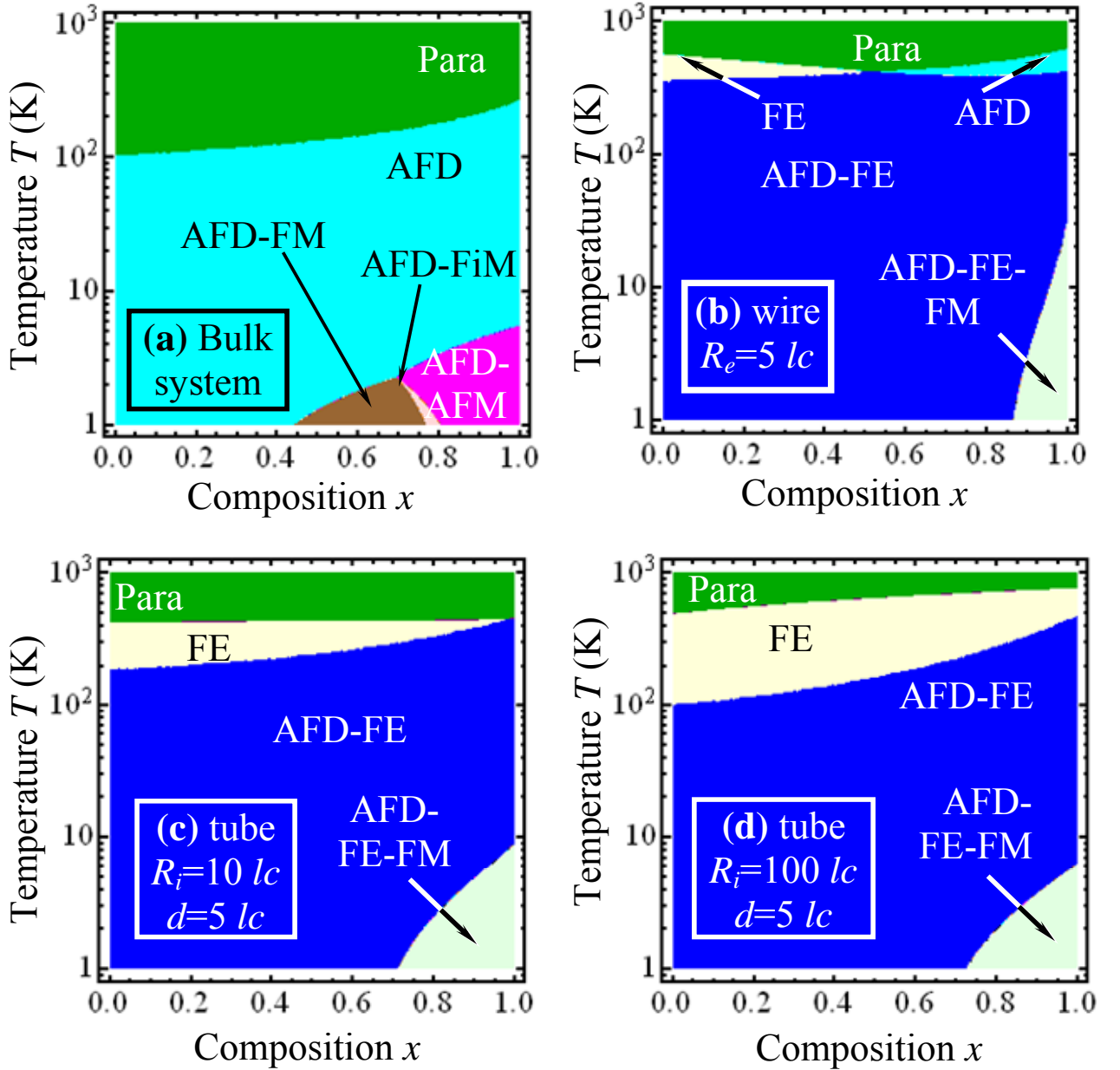

**Figure 2.** Predicted temperature-composition phase diagrams of **(a)** bulk $Eu_xSr_{1-x}TiO_3$ system, where Para (paraelectric-paramagnetic), AFD (antiferrodistortive), AFD-FM (antiferrodistortive-ferromagnetic), AFD-AFM (antiferrodistortive-antiferromagnetic), and AFD-FiM (antiferrodistortive-ferrimagnetic) phases are present, **(b)** $Eu_xSr_{1-x}TiO_3$ wire of radius 5 *lc*, **(c)** $Eu_xSr_{1-x}TiO_3$ nanotube of radius 10 *lc*, and **(d)** $Eu_xSr_{1-x}TiO_3$ nanotube of radius 100 *lc*. The tubes of thickness for (c) and (d) are 5 *lc* with a tensile misfit strain $u_c = +3\%$. The surface tension coefficient $\mu = 30$ N/m for nanowire and nanotubes. The existing phases in the nanosystems are Para, FE, AFD, AFD-FE, and AFD-FE-FM.

Analyzing the phase diagrams predicted for the bulk systems (**Fig. 2a**), we observed the unexpected appearance of the ferromagnetic ordering (AFD-FM and AFD-FiM phases) for Eu content $x > 0.4$. This FM ordering is induced by the substitution of Sr atoms at the Eu sites. Therefore, we predict that nonmagnetic $Sr^{+2}$ ions could induce magnetization due to the indirect super-exchange between Eu atoms via Sr atoms rather than via oxygens, namely Sr-diluted ferromagnetism for composition $x$ from 0.45 to 0.75 or ferrimagnetism for $x \approx 0.8$. On the other hand FM ordering may originate from spin canting [63, 64], especially if the energies of different magnetic orderings (A-, C-, F-, and G-types) are very close. Therefore, bulk solid solution $Eu_xSr_{1-x}TiO_3$ should be included to the multiferroic family. Note that the crossover of the AFD magnetic phases AFD-AFM → AFD-FiM → AFD-FM originated from the magnetic percolation model, namely due to the different percolation thresholds for ferromagnetizm $x_{cr}^F \approx 0.24$ and antiferromagnetizm $x_{cr}^A \approx 0.48$, which is in agreement with classical percolation theory [47]. We hope that this prediction will be verified either experimentally or from the first-principles calculations.

For $Eu_xSr_{1-x}TiO_3$ nanotubes and nanowires, our calculations demonstrated that several ordered phases can be thermodynamically stable under tensile strain (see **Figs. 2b-d**), namely Para, FE, AFD, AFD-FE, and AFD-FE-FM. Note, that ferroelectric FE, AFD-FE, and AFD-FE-FM phases are absent in the bulk $Eu_xSr_{1-x}TiO_3$, since ferroelectric ordering appearance in incipient ferroelectrics is possible for small sizes only [15-25].

In accordance with our calculations, general conditions of the multiferroic AFD-FE-FM stability in nanosystems are (i) small thickness and radius, (ii) relatively low temperatures (< 20 K), (iii) high Eu content ($x > 0.7$), and (iv) positive tensile misfit strains $u_c > 0$. For instance, the temperature – composition phase diagram of $Eu_xSr_{1-x}TiO_3$ nanowires, calculated for tensile strains $u_c = +3\%$ (**Fig. 2b**), demonstrates that magnetoelectric AFD-FE-FM phase appears in the nanowires for Eu content $x$ more than $x_c \approx 0.85$ and temperatures less than 20 K. **Figures 2c-d** illustrate that AFD-FE-FM phase appears in $Eu_xSr_{1-x}TiO_3$ nanotubes for Eu concentration more than $x_c \approx 0.7$, and temperatures lower than 10 K. Increase of the Eu composition from $x_c$ to 1 essentially enlarges the temperature interval of AFD-FE-FM phase stability. The diagram of $Eu_xSr_{1-x}TiO_3$ nanowires contains smaller $x$-region of AFD-FE-FM phase (namely $0.85 \leq x \leq 1$, see **Fig. 2b**) in comparison to the corresponding $x$-region of nanotubes ($0.7 \leq x \leq 1$, see **Figs. 2c-d**). This is because of the additional contribution of misfit-related strain terms $\sim u_c \left( R_i^2 / R_e^2 \right)$ (Eqs. 5) originating from the tube-core lattice constants mismatch for nanotubes. **Figure 2d** presents the limit of the thin epitaxial $Eu_xSr_{1-x}TiO_3$ film with in-plane polarization, where the ADF-FE-FM phase appeared from misfit effect, corresponding term is $\dfrac{Q_{11}(x)+Q_{12}(x)}{s_{11}(x)+s_{12}(x)} \dfrac{R_i^2}{R_e^2} u_c \approx \dfrac{Q_{11}(x)+Q_{12}(x)}{s_{11}(x)+s_{12}(x)} u_c$.

Therefore, it is important to emphasize that the misfit strain existing between the nanotube-core interface allows the possibility of controlling the phase diagram of the $Eu_xSr_{1-x}TiO_3$ nanotubes. Such possibility is absent for nanowires. The misfit strain - composition phase diagrams of $Eu_xSr_{1-x}TiO_3$ nanotubes with internal radius 10 *lc*, outer radius 15 *lc*, and thickness *d*=5 *lc* are shown in **Fig. 3** at two different temperatures, at 4 K (low temperature), and at 300 K (room temperature). From **Fig. 3a** it is clear that the FM properties of $Eu_xSr_{1-x}TiO_3$ nanotubes can appear at about $x_c > 0.8$, which is much higher than the percolation threshold of $x_{cr}^F \approx 0.24$ at low temperatures $T < 4$ K and positive tensile strains ($u_c > 0$). The region of AFD-FE-AFM stability becomes narrower with the increase in temperature and it disappears at higher temperatures. At low temperatures (≤ 10 K) there are two stable multiferroic phases, namely AFD-FE-AFM and AFD-FE-FM. Additional calculations (data not shown) proved that only pure $EuTiO_3$ can be AFM for given sizes and strains at $T > 10$ K. At room temperature (**Fig. 3b**) the disordered Para phase appears at $x < 0.8$ and $u_c < +1\%$. Such enlarged region of the para phase occurs at room temperature because of the absence of the axial ferroelectric polarization $P_3$ in the compressed nanotubes (similar effect is reported for compressed ferroelectric films [53]). Due to the strong depolarization field $E_\rho^d \sim -P_\rho/\varepsilon_0\varepsilon_b$, the ferroelectric phase with radial polarization $P_\rho$ perpendicular to the surface of the tube/wire may appear at very high compressive strains $u_c < -5$ % [65].

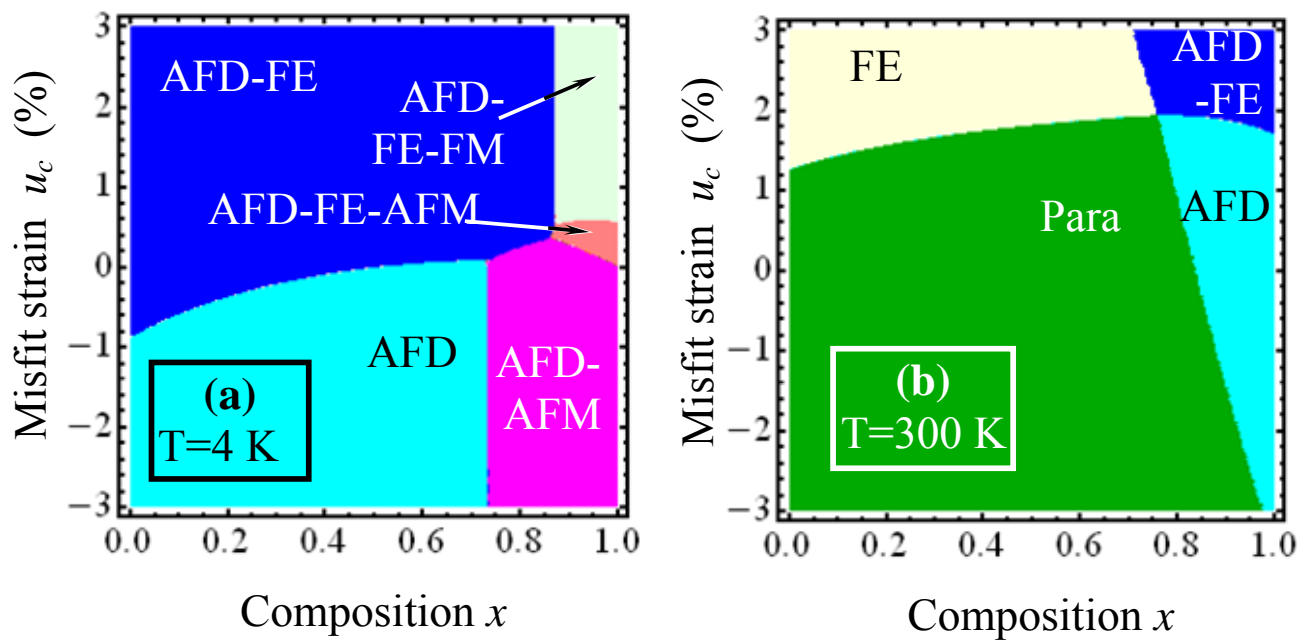

**Figure 3.** The misfit strain–composition phase diagrams of $Eu_xSr_{1-x}TiO_3$ nanotube with internal radius 10 *lc*, outer radius 15 *lc*, and thickness *d*=5 *lc* at **(a)** temperature *T* = 4 K, and **(b)** at *T* = 300 K.

In addition, the phase diagram in coordinates of misfit strain–temperature for different Eu compositions (*x* = 0, 0.5, 0.75 and 1) are presented for $Eu_xSr_{1-x}TiO_3$ nanotubes with internal radius 10 *lc*, outer radius 15 $lc$, and thickness $d=5$ $lc$ (**Fig. 4**). The results show the transition of different phases where 4 phases are present for $x \leq 0.5$ ((Para, FE AFD, and AFD-FE), which transforms to 8 phases for $x = 0.75$ (Para, AFD, FE, AFD-FE, AFD-AFM, AFD-FiM, AFD-FM and AFD-FE-FM) and then to 6 phases for $x = 1$ (Para, AFD, AFD-FE AFD-AFM, AFD-FE-AFM and AFD-FE-FM). It is evident from **Fig. 4** that the phase boundaries have relatively small horizontal slopes (i.e. these are weakly misfit-dependent), while the content of Eu influences the vertical position of the phase boundaries. These results are consistent with the model assumptions, as the model did not consider the gradient effects and stress relaxation via the appearance of dislocations.

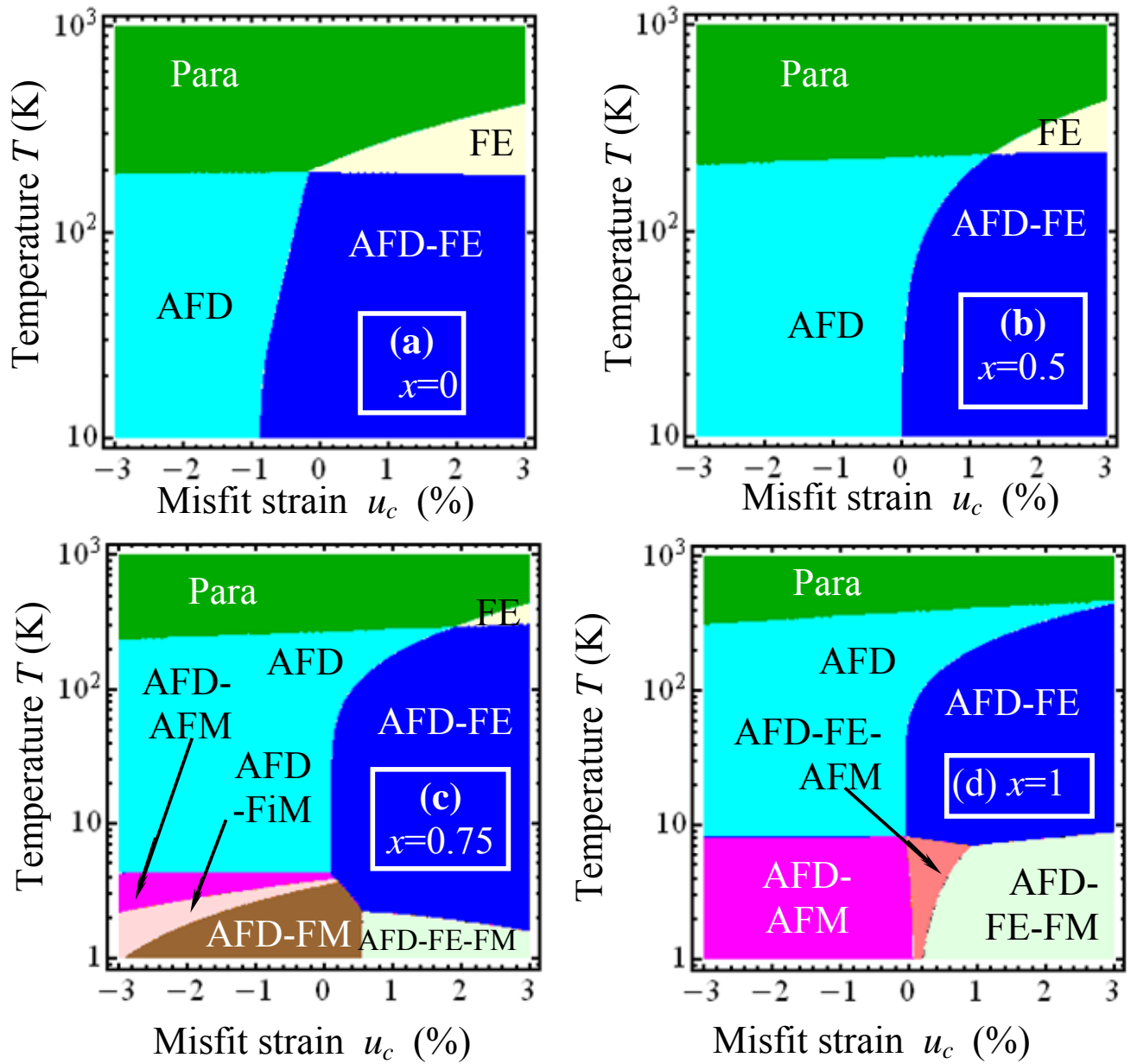

**Figure 4.** Temperature - misfit strain phase diagrams of $Eu_xSr_{1-x}TiO_3$ tube with internal radius 10 $lc$, outer radius 15 $lc$ and thickness $d=5$ $lc$ for different Eu compositions **(a)** $x=0$, **(b)** $x = 0.5$, **(c)** $x = 0.75$, and **(d)** $x = 1$.

From **Figs. 3** and **4**, it is important to emphasize that the multiferroic ADF-FE-FM phase in $Eu_xSr_{1-x}TiO_3$ nanosystems can be stable only for tensile strain ($u_c > 0$). The multiferroic phases are absent for zero strain ($u_c = 0$) and compressive strains ($u_c < 0$). This is because the FE phases with spontaneous polarization $P_3$ parallel to the tube axis become unstable for zero ($u_c = 0$) and negative ($u_c < 0$) strains. The phases with spontaneous polarization $P_\rho$ perpendicular to the tube surface $\rho = R_e$ would become stable, however they appeared to be completely suppressed by the strong depolarization field $E_\rho^d \sim -P_\rho/\varepsilon_0\varepsilon_b$, since we did not impose any type of short-circuit conditions at the tube/wire sidewalls. The effect of tensile strains can be readily explained from Eqs.(5a-b), where the terms, $\frac{Q_{11}+Q_{12}}{s_{11}+s_{12}}\frac{R_i^2}{R_e^2}u_c$ and $\frac{Z_{11}+Z_{12}}{s_{11}+s_{12}}\frac{R_i^2}{R_e^2}u_c$, should become positive in order to increase effective FE and FM Curie temperatures. Since $\frac{Q_{11}+Q_{12}}{s_{11}+s_{12}} > 0$ and $\frac{Z_{11}+Z_{12}}{s_{11}+s_{12}} > 0$ in perovskites, the terms are positive under the condition $u_c > 0$.

In general we predict that tensile misfit strains are necessary for the appearance of multiferroic phase in $Eu_xSr_{1-x}TiO_3$ nanosystems. Similar to tensile misfit strains, high positive surface tension coefficients ($\mu$) increase the effective Curie temperatures for both nanotubes and nanowires [15]. The effect of surface tension can be readily explained from Eqs.(5a), since for positive surface tension coefficient $\mu$ and negative electrostriction coefficient $Q_{12}$, the term $4Q_{12}\mu/R_e$ increases the effective Curie temperatures for both nanotubes and nanowires.

We have also estimated the effect of wire radii and tube thickness on the phase diagram of $Eu_xSr_{1-x}TiO_3$ nanosystems. **Figure 5** illustrates the phase diagrams with respect to temperature-nanowire radius (**Fig. 5a**) and temperature-tube thickness (**Figs. 5b-c**) for composition $x = 0.9$ and misfit strain $u_c = +3\%$. Since the multiferroic phases are present at low temperatures, only $T < 10$ K are evaluated. For the nanowire system, the multiferroic AFD-FE-FM phase (light green region) is stable below 3 K (**Fig. 5a**). Similar analysis on the nanotube systems show that the multiferroic phase is stable at $T < 6$ K for smaller nanotubes (with $R_i = 10$ $lc$ (**Fig. 5b**)), and at $T < 4$ K for larger nanotubes (with $R_i = 100$ $lc$ (**Fig. 5c**)). Comparing the nanowire and nanotube systems, the temperature stability of the multiferroic phases is higher for the nanotube systems. This difference in temperature stability can be attributed to the misfit strains present in nanotube systems. The $Eu_xSr_{1-x}TiO_3$ tube with inner radius of 100 $lc$ virtually presents the limit of the thin epitaxial film with in-plane polarization, where the AFD-FE-FM phase appeared from misfit effect only. This independence of the AFD-FE and AFD-FE-FM phase boundary on the tube thickness in **Fig. 5c** is due to the lack of misfit dislocation and polarization gradient effect consideration in the current model.

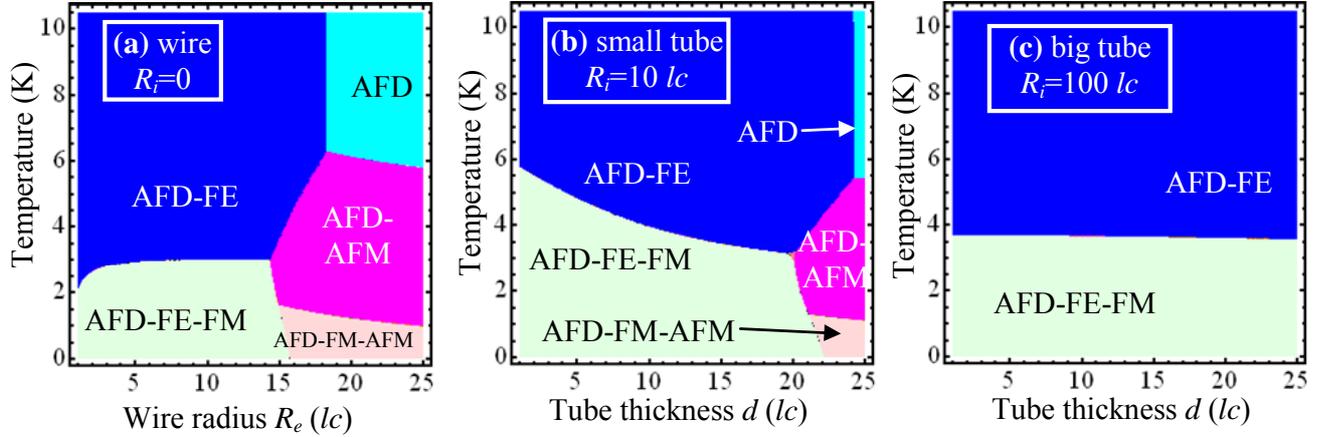

**Figure 5.** Phase diagrams of $Eu_xSr_{1-x}TiO_3$ **(a)** nanowire with $R_i = 0$, **(b)** nanotube with $R_i = 10\ lc$ and **(c)** nanotube with $R_i = 100\ lc$ in coordinates wire radius $R_e$ or tube thickness $d$ vs. temperature $T$ calculated for composition $x=0.9$ and misfit strain $u_c=+3\%$.. Other parameters and phase designations are the same as in **Fig. 2**.

## IV. SPONTANEOUS POLARIZATION AND MAGNETIZATION

The spontaneous polarization and magnetization vs. composition $x$ of Eu in $Eu_xSr_{1-x}TiO_3$ nanowires and nanotubes are shown in **Fig. 6** for fixed radii, tensile misfit strain and different temperatures (specified near the individual curves). Note that spontaneous magnetization is absent at compressive strains and thus the case with tensile misfit strain of +3% is considered. It is observed that spontaneous polarization increases with the reduction in temperature. The magnitude of spontaneous polarization increases with the increase in Eu content for most of the temperatures. However, the trend is not followed above 280 K, which is the temperature of the structural phase transition in bulk $EuTiO_3$ (~280 K). Spontaneous magnetization abruptly appears with Eu content more than the threshold value $x_c$ and at temperatures less than 10 K. Such abrupt composition-induced FM phase transition is of the first order. It is clear from **Fig. 6** that FE phase exists at all $x$ and temperatures less than 300 K. The jumps on spontaneous polarization values at low temperatures (4 K data in **Figs. 6a-b**) matches with the simultaneous appearance of spontaneous magnetization phases, i.e. they indicate magnetoelectric FE-FM phase transition.

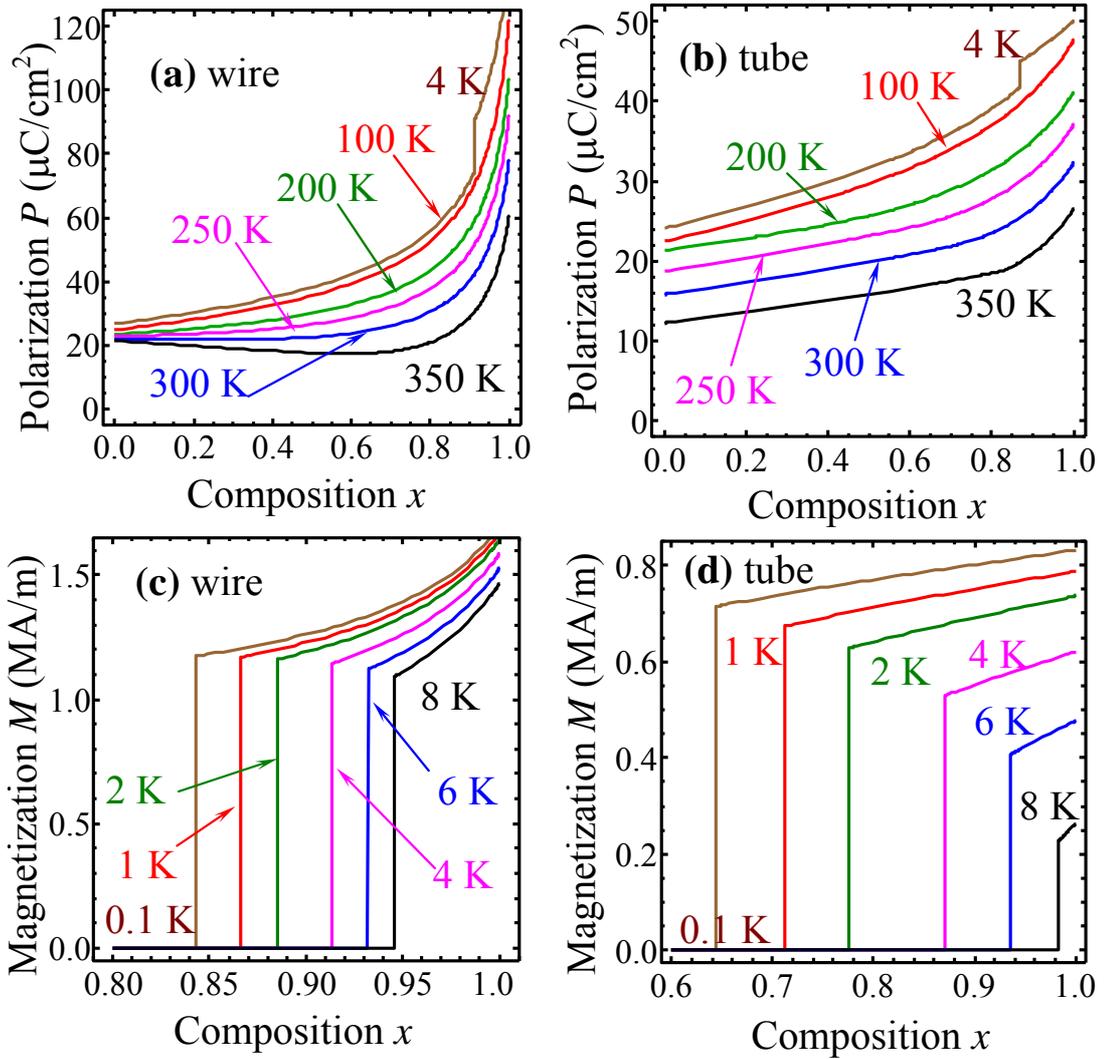

**Figure 6.** Change in spontaneous polarization vs. composition $x$ of $Eu_xSr_{1-x}TiO_3$ **(a)** nanowire, and **(b)** nanotube at different temperatures. Change in spontaneous magnetization vs. composition $x$ of $Eu_xSr_{1-x}TiO_3$ **(c)** nanowire, and **(d)** nanotube at different temperatures. The nanowire is of radius 5 $lc$ while the nanotube is with internal radius 10 $lc$ and thickness 5 $lc$. The results are shown for tensile misfit strain of +3%. The temperature values are specified near the curves.

Spontaneous polarization and magnetization dependence on the wire radii or the tube thickness of $Eu_xSr_{1-x}TiO_3$ nanosystem are shown in **Fig. 7**. The calculations are performed for Eu content $x = 0.9$, tensile misfit strain +3% and at different temperatures. For wire radius less than 10 $lc$, the spontaneous polarization reaches rather high values ~20 – 100 μC/cm² up to room temperatures (**Fig. 7a**). At low temperatures (<10 K) and small tube thickness (<10 $lc$), spontaneous polarization reaches a maximum value of ~50 μC/cm² (**Fig. 7b**). Note, that spontaneous polarization increases with the reduction in wire radii or tube thickness. It is seen from the plots that FE polarization and FM magnetization disappear when the tube thickness overcomes the critical value. The tube critical thickness for spontaneous polarization disappearance is temperature dependent; it decreases with the increase in temperature from 20 $lc$ at 3 K to 7 $lc$ at 300 K (**Fig. 7b**). The quantitative analysis of spontaneous magnetizations at 3 K are characterized to be ~0.7 MA/m for nanowires of radii range 8 $lc < R_e <$ 15 $lc$ (see **Fig. 7c**), while ~0.5 MA/m for nanotubes of thickness less than 20 $lc$ (see **Fig. 7d**). These results are in agreement with the results presented in **Figs. 5a** and **b**. For nanotubes the tube-on-core geometry seems more favorable for ferromagnetism.

It is seen from **Fig. 7** that the size-induced ferroelectric phase transition is of the second order at 300 K and of the first order at 3 K, while the size-induced ferromagnetic phase transition is of the first order at 3 K. Numerical simulations proved that all the first order size-induced ferroelectric phase transitions are the transitions between two AFM phases (e.g. AFD-AFM and AFD-FE-AFM), while, the second order transitions correspond to Para-FE transition. All jumps at low temperatures indicate magnetoelectric FE-FM transitions.

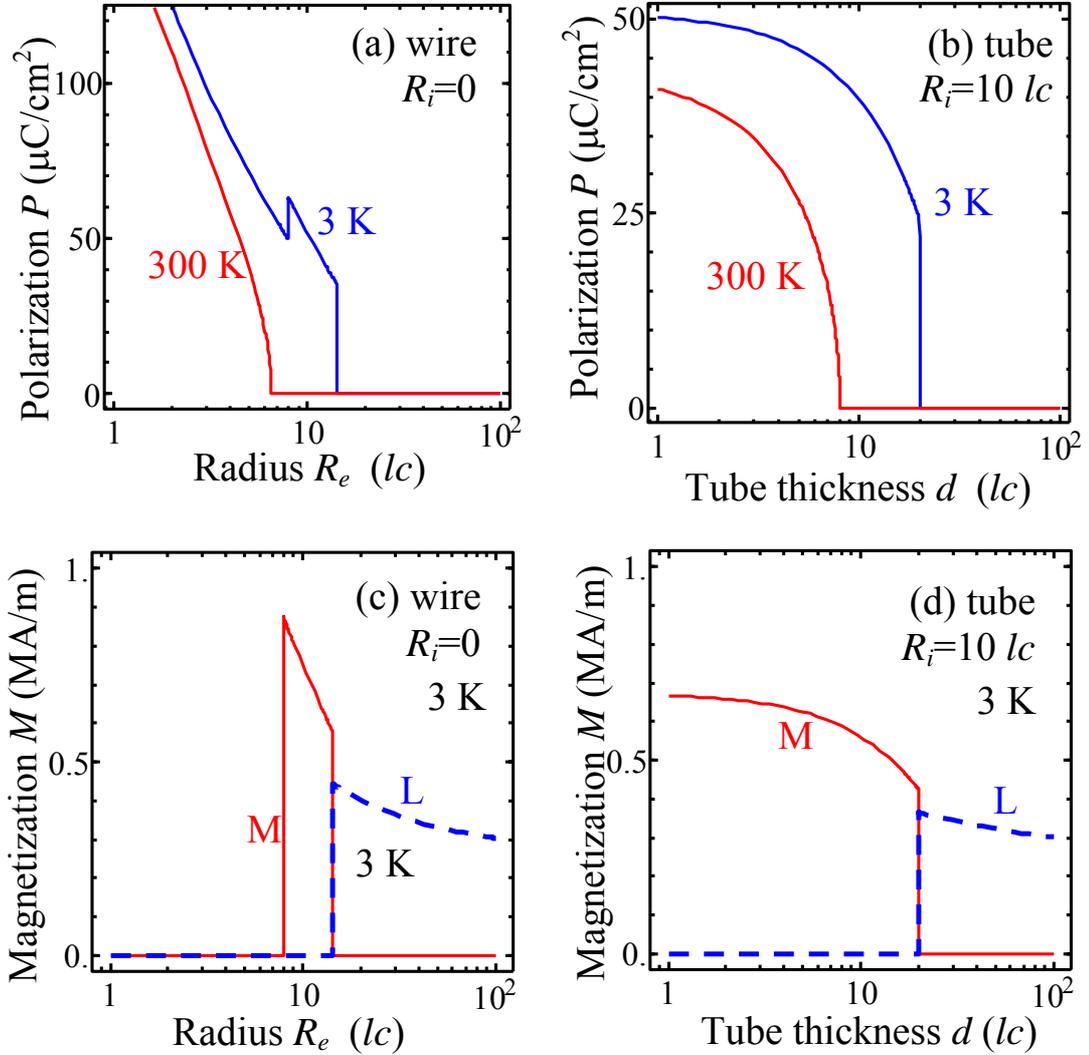

**Figure 7.** Change in spontaneous polarization with respect to **(a)** wire radii, and **(b)** tube thickness in $Eu_xSr_{1-x}TiO_3$ nanosystem calculated for $x = 0.9$, and tensile misfit strain $u_c = +3\%$. Change in magnetization M and antimagnetization L with respect to **(c)** wire radii, and **(d)** tube thickness. Polarization is shown for two different temperatures, 3 K and 300 K.

## V. DISCUSSION

In this study we have calculated the phase diagrams of bulk $Eu_xSr_{1-x}TiO_3$, and $Eu_xSr_{1-x}TiO_3$ nanosystems (nanotubes, and nanowires) using LGD theory. For bulk $Eu_xSr_{1-x}TiO_3$ solid solution, the FM phase is predicted to be stable at low temperatures. The presence of FM phase due to the substitution of a nonmagnetic ion may be explained by the super-exchange mechanism. It is reported that in bulk $EuTiO_3$ AFM phase can originate at $T < 5.5$ K from direct AFM type exchange interaction between $Eu^{+2}$ ions and indirect super-exchange $Eu^{2+} – O^{2-} – Eu^{2+}$ [12]. Therefore, it may be assumed that a super-exchange between $Eu^{2+} – Sr^{+2} – Eu^{2+}$ could be responsible for the ferromagnetic phase appearance in bulk $Eu_xSr_{1-x}TiO_3$ solid solution. Similar phenomenon of nonmagnetic $Ba^{+2}$ substitution on magnetization is reported for $Pb_{1-x}Ba_xFe_{1/2}Nb_{1/2}O_3$ by Raevski et al. [66]

From our calculations, nanosized $Eu_xSr_{1-x}TiO_3$ wires and tubes under favorable conditions can be multiferroics. Magnetoelectric AFD-FE-FM phase, which is the most important phase for multiferroic applications, can exist at Eu content more than critical ($x_c \approx 0.75$), at tensile strains ($u_c \geq +1\%$) and low temperatures < 10 K. Increase of the Eu composition $x$ from $x_c$ to 1 essentially enlarges the region of ADF-FE-FM phase stability at a fixed temperature. The maximal polarization (25 – 100 $\mu C/cm^2$) and magnetization (0.5 - 1 MA/m) values (see **Fig. 8**) are high or comparable with proper ferroelectrics $BaTiO_3$ (26 $\mu C/cm^2$) [56], $LiNbO_3$ (75 $\mu C/cm^2$) [56], $PbTiO_3$ (81 $\mu C/cm^2$) [67] and typical ferromagnetics [68]. The existence of ME coupling manifests itself by jumps in maximum polarization (**Fig. 8a**) at the content $x$ of magnetization appearance (**Fig. 8b**). The temperature region of magnetization existence depends on the size of the nanosystems and the Eu composition $x$. Therefore, the appearance of the multiferroic phase depend on the choice of the amount of substitution ($x$), the rigid core material for misfit strain ($u_c$), the radius and thickness of the nanosystems, and temperature.

All results presented in this article for $Eu_xSr_{1-x}TiO_3$ are based on the parameters of $EuTiO_3$ and $SrTiO_3$ taken from experiments [9, 14], independent first principles calculations [13] and our first principles calculations (see **Tables 1, 2** and **Suppl.Mat**). Although the sources we collected the parameters are quite reliable, the collective use of different parameter sets may result in certain uncertainties in the prediction. For example, the physical origin of some of the predicted phenomena, such as the increase of FE phase transition temperature with the increase in Eu content in nanosystems and the appearance of FM phase with dilution of bulk $EuTiO_3$ by $SrTiO_3$ require further investigation and experimental verification.

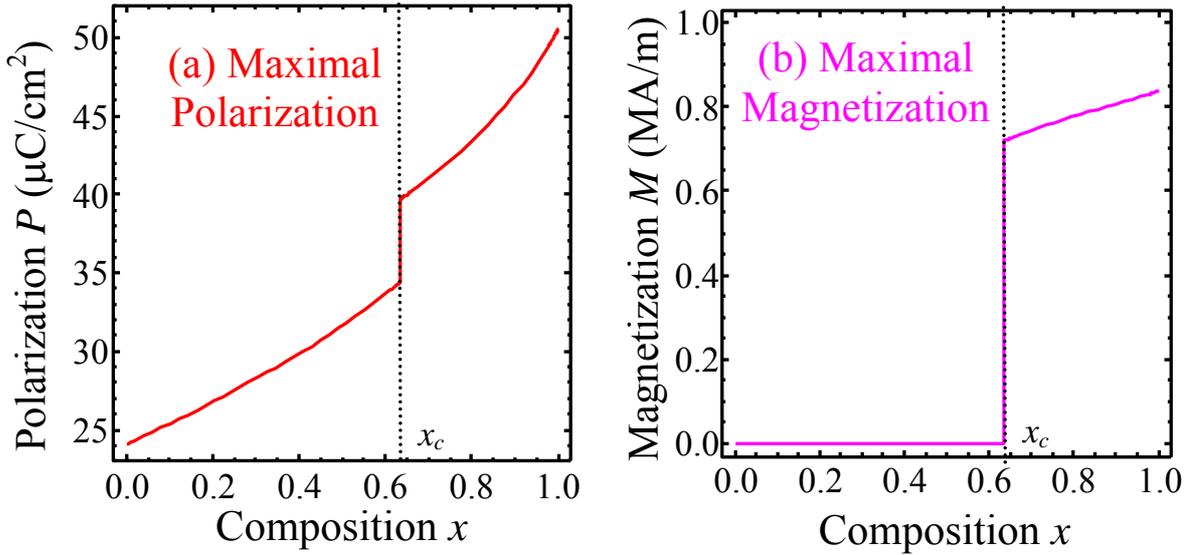

**Figure 8.** The **(a)** maximal polarization, and **(b)** maximal magnetization values calculated for $Eu_xSr_{1-x}TiO_3$ nanotube of internal radius 10 $lc$, thickness 5 $lc$, for +3% tensile strain at zero Kelvin. Antimagnetization $L$ is absent for the chosen parameters.

We hope that our predictions will stimulate experimental and computational studies of $Eu_xSr_{1-x}TiO_3$ nanosystems, where the coupling between structural distortions, polarization and magnetization can lead to the versatility and tenability of the magnetoelectric multiferroic phases.

**Acknowledgements.** E. A. E, M. D.G. and A. N. M. acknowledge Science and Technology Center of Ukraine, project STCU-5514.

# SUPPLEMENT

to

# NEW MULTIFERROICS BASED ON Eu$_x$Sr$_{1-x}$TiO$_3$ NANOTUBES AND NANOWIRES

by

Eugene A. Eliseev, Maya D. Glinchuk, Victoria V. Khist, Chan-Woo Lee, Chaitanya S. Deo,

Rakesh K. Behera, and Anna N. Morozovska

In this supplement we have discussed the fitting procedures used to estimate various coefficients for tilt, polarization and magnetization components of the LGD free-energy expressions (Eqs. 1-5 in the main text). Since the parameters related to SrTiO$_3$ are well established in the literature, all the discussions in this supplement are related the parameters for EuTiO$_3$ only.

**Part –I: Determination of tilt related parameters for EuTiO$_3$:**

First we have considered the free energy dependence on the antiferrodistortive (AFD) order parameter $\Phi$:

$$\Delta G_\Phi = \frac{\alpha_\Phi}{2}\Phi^2 + \frac{\beta_\Phi}{4}\Phi^4 - (R_{11}\sigma_{11} + R_{12}\sigma_{22} + R_{12}\sigma_{33})\Phi^2 \tag{S.1a}$$

Here $\alpha_\Phi$ is temperature dependent, while $\beta_\Phi$ is usually constant, $\sigma_{ij}$ is the elastic stress tensor, $R_{ij}$ is the rotostriction coefficient. The order parameter $\Phi$ could be measured as either tilt angle of oxygen octahedra or oxygen displacements from the symmetric positions in ideal perovskite structure. The spontaneous value of the tilt is

$$\Phi_S = \sqrt{-\alpha_\Phi/\beta_\Phi}. \tag{S.1b}$$

and the corresponding free energy for mechanically free system is

$$\Delta G_\Phi\big|_{\sigma=0} = \frac{-\alpha_\Phi^2}{4\beta_\Phi}. \tag{S.1c}$$

Using the temperature dependence of oxygen octahedra tilt angle $\phi$ from Allieta et al. [1], we recalculated the oxygen displacement $\Phi = \frac{a}{2}\tan(\phi)$ (see **Fig. S1**) with experimentally observed lattice constant $a$. We fitted the temperature–tilt angle plot by substituting the temperature dependence of the expansion coefficient $\alpha_\Phi$ in the form of Barret equation in Eq. S.1b, where $\alpha_\Phi = \alpha_T^{(\Phi)} \frac{T_q^{(\Phi)}}{2}\left(\coth\left(\frac{T_q^{(\Phi)}}{2T}\right) - \coth\left(\frac{T_q^{(\Phi)}}{2T_c^{(\Phi)}}\right)\right)$. This fitting procedure provides a combined influence of the $\alpha_T^{(\Phi)}$ and $\beta_\Phi$ parameters. Thus we have obtained the combination

$$\sqrt{\frac{\alpha_T^{(\Phi)}}{\beta_\Phi}\frac{T_q^{(\Phi)}}{2}\coth\left(\frac{T_q^{(\Phi)}}{2T_c^{(\Phi)}}\right)} \equiv \Phi_q = 23 \text{ pm}$$, and temperatures $T_c^{(\Phi)} = 220$ K and $T_q^{(\Phi)} = 205$ K from the fitting procedure.

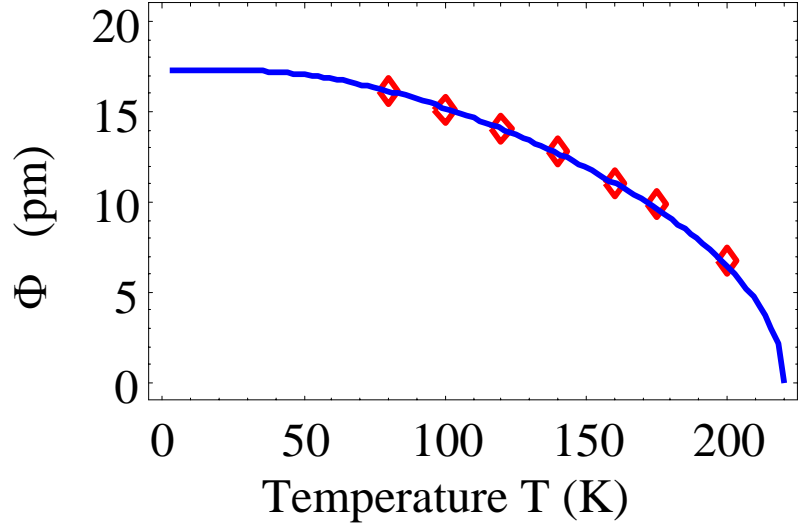

**Figure S1.** Relationship between temperature and antiferrodistortive (AFD) order parameter in EuTiO$_3$. In the plot, open symbols represent experimental data of Allieta et al., while the solid line represents the fitting performed with $\Phi_q = 23$ pm and $T_c^{(\Phi)} = 220$ K and $T_q^{(\Phi)} = 205$ K.

In order to derive the individual values $\alpha_T^{(\Phi)}$ and $\beta_\Phi$, we have used independent measurements of specific heat, which is given as $\Delta C = -T\,\partial^2 \Delta G/\partial T^2$. Using Eq.S.1c, we have derived the expression of specific heat in terms of $\alpha_\Phi$ and $\beta_\Phi$ as:

$$\Delta C = T\frac{1}{2\beta_\Phi}\left(\frac{\partial \alpha_\Phi}{\partial T}\right)^2 = T\frac{1}{2\beta_\Phi}\frac{\left(-\alpha_T^{(\Phi)}\right)^2}{\left(\frac{2T}{T_q^{(\Phi)}}\sinh\left(\frac{T_q^{(\Phi)}}{2T}\right)\right)^4}, \quad T < T_c^{(\Phi)} \tag{S.2}$$

**Figure S2** represents the temperature dependence of specific heat in EuTiO$_3$ obtained by Bussmann-Holder et al. [2]. The results show a ~ 3 J/(mol K) drop in specific heat around the Curie temperature. Therefore considering $\Delta C \approx 3$ J/(mol K) at the Curie temperature, $\sqrt{\frac{\alpha_T^{(\Phi)}}{\beta_\Phi}\frac{T_q^{(\Phi)}}{2}\coth\left(\frac{T_q^{(\Phi)}}{2T_c^{(\Phi)}}\right)} = 23$ pm, $T_c^{(\Phi)} = 220$ K, $T_q^{(\Phi)} = 205$ K, and using Eq.(S.2), we have estimated the values of $\alpha_T^{(\Phi)}$ and $\beta_\Phi$ as:

$$\alpha_T^{(\Phi)} = 3.913 \times 10^{26} \text{ J/(m}^5 \text{ K)}, \text{ and } \beta_\Phi = 1.744 \times 10^{50} \text{ J/m}^7. \tag{S.3}$$

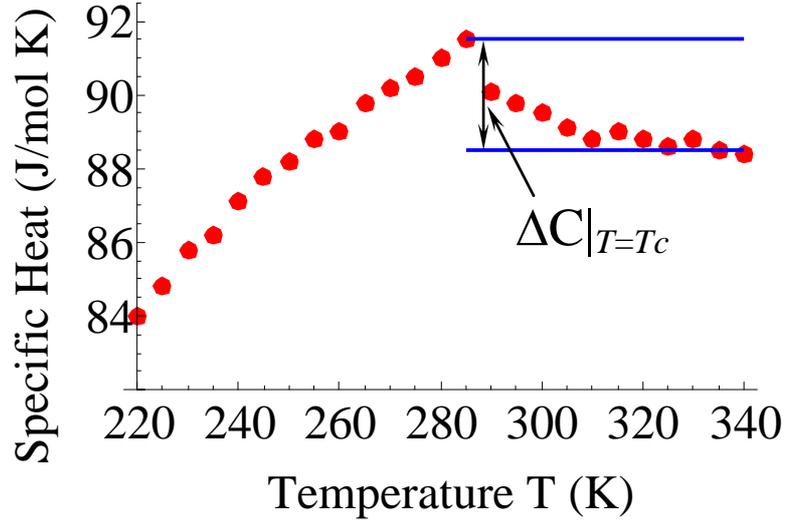

**Figure S2.** Temperature dependence of specific heat in $EuTiO_3$ (the filled circles represent experimental data of Bussmann-Holder et al.[2]). Definition of the drop in specific heat at the transition temperature is shown by solid lines.

After calculating $\alpha_T^{(\Phi)}$ and $\beta_\Phi$, we have estimated the values of rotostriction coefficients $R_{11}$ and $R_{12}$ which rrelate spontaneous strain in the tetragonal phase with the square order parameter (tilt). In order to calculate the $R_{ij}$ values, we have considered the lattice variation with temperature for $EuTiO_3$. Following the work by Allieta et al. [1], we extrapolated the cubic lattice parameter using $a_o = 3.90478(1+9.33\times10^{-6}(T-293))$ Å. **Figure S3** represents the experimental lattice parameter variation with temperature and the linear extrapolation used for fitting $R_{11}$ and $R_{12}$. Then we fitted the lattice constants with the following relationships:

$$a = a_0(1+u_{11}), \qquad c = a_0(1+u_{33}) \qquad (S.4a)$$

$$\text{where } u_{11} = R_{12}\Phi_S^2 \text{ and } u_{33} = R_{11}\Phi_S^2 \qquad (S.4b)$$

Using the spontaneous value of the tilt $\Phi_S$ and $a_0$, our fitting procedure for spontaneous strain components ($u_{11}$ and $u_{33}$ are shown in **Fig. S4**) resulted in

$$R_{11}=5.46 \times 10^{18} \text{ m}^{-2} \text{ and } R_{12} = -2.35 \times 10^{18} \text{ m}^{-2} \qquad (S.5)$$

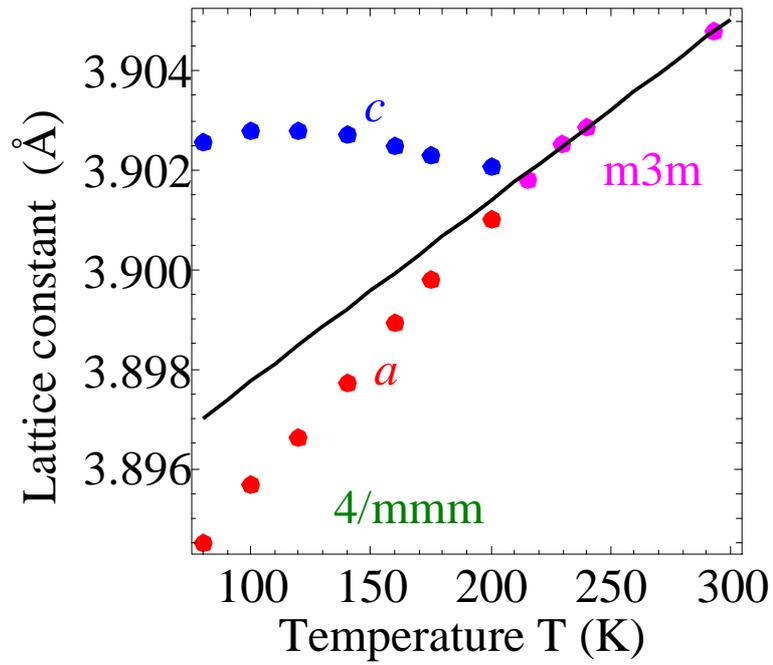

**Figure S3.** Illustration of change in lattice parameter with temperature in $EuTiO_3$. (The filled circles represent experimental data from Allieta et al. [1], and the solid line is for the linear extrapolation of the cubic lattice constant $a_0 = 3.90478(1+9.33\times10^{-6}(T-293))$ Å.

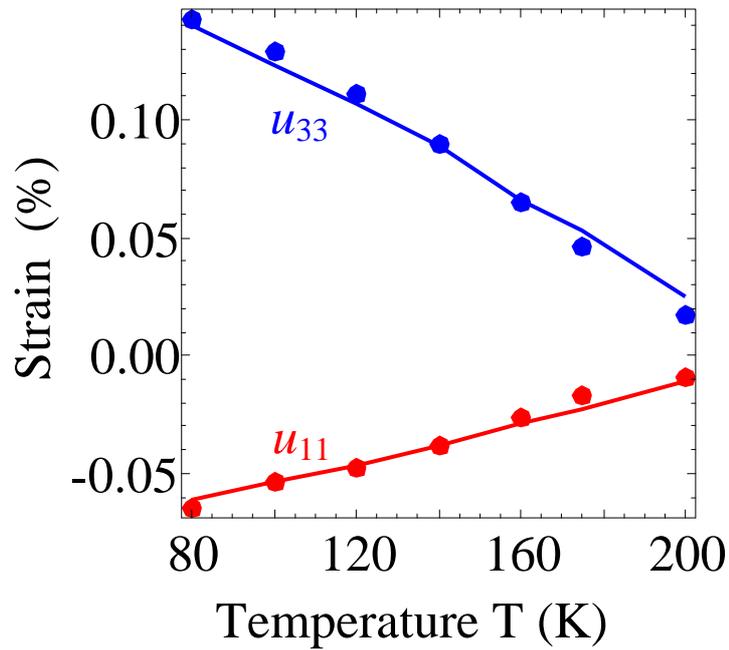

**Figure S4.** Temperature dependence of spontaneous strain in $EuTiO_3$. The filled circles represent recalculated experimental data from Allieta et al.[1] **(Fig. S3)**, and the solid lines represent the fitted data.

**Part –II: Determination of polarization and magnetization related parameters for $EuTiO_3$:**
Katsufuji and Takagi [3] reported the temperature and magnetic field dependence of dielectric permittivity ε and magnetic susceptibility of EuTiO$_3$ in the range of 2 K -100 K and magnetic fields from 0 to 5 Tesla. From these data they deduced the temperature dependence of $\alpha_P$ (Barret equation) and $\alpha_M = \alpha_{MT}(T - T_C)$ corresponding to the bulk para-phase. V. Goian et al. [4] obtained temperature dependence static magnetic permittivity for the paramagnetic and antiferromagnetic phases using the infrared reflectivity spectra analysis over a wide range of temperature.

Katsufuji and Takagi [3] also fitted strong changes in ε in the antiferromagnetic (AFM) phase of EuTiO$_3$ below 5.5 K using average spin calculated from the mean-field model. However, the fitting did not consider the emergence of structural order parameter (spontaneous tilt of oxygen octahedra), which is reported recently between 200 K - 300 K [1, 2]. Thus we modified the phenomenological model to include the tilt contribution at higher temperatures. Therefore, the updated dielectric permittivity, which includes the multiferroic AFM-AFD phase is given as:

$$\varepsilon = \varepsilon_b + \frac{1}{\varepsilon_0 \left( \alpha_P(T) + \eta_{P\Phi}\Phi^2 + \gamma_{AFM} L^2 \right)}. \tag{S.6}$$

Here $\alpha_P = \alpha_T^{(P)} \frac{T_q^{(P)}}{2} \left( \coth\left(\frac{T_q^{(P)}}{2T}\right) - \coth\left(\frac{T_q^{(P)}}{2T_c^{(P)}}\right) \right)$, and the order parameter for the AFM phase is represented as:

$$L = \sqrt{-\frac{\alpha_L}{\beta_L}}. \tag{S.7}$$

Following the work of Katsufuji and Takagi [3], we could assume that for T→0 the spins of the magnetic sublattices are saturated to 7/2 per unit cell (but have alternating signs). Hence the value of L is calculated as half the sum of the magnetization of the sublattices, which results in $L = \frac{7}{2}\frac{\mu_B}{a_0^3} = 0.509 \times 10^6$ A/m. We assume that the temperature coefficient $\alpha_N = 2\pi \cdot 10^{-6}$ Henri/(m·K) is the same for $\alpha_M$ and $\alpha_L$, i.e. $\alpha_M = \alpha_N(T - T_C)$ and $\alpha_L = \alpha_N(T - T_N)$. Temperatures $T_C$ = 3.5 K and $T_N$ = 5.5 K are the Curie and Néel temperatures of EuTiO$_3$ respectively [3, 5], we could estimate $\beta_L = 1.3 \times 10^{-16}$ J m/A$^4$ using Eq.S.7.

Therefore, using L and the magnitudes of $\alpha_T^{(\Phi)}$ and $\beta_\Phi$ obtained in Eq.(S.3) for the tilt parameters, we have fitted the experimental results on the temperature variation of dielectric constant in EuTiO$_3$. The fitting results are shown in **Fig. S5**.

This fitting procedure resulted in the following values of parameters:

$$\alpha_T^{(P)} = 1.95 \times 10^6 \text{ m/(F·K)}, \; T_q^{(P)} = 230 \text{ K}, \; T_c^{(P)} = -133.5 \text{ K} \tag{S.8a}$$

$$\eta_{P\Phi} = -4.45\times 10^{29}\,\text{F}^{-1}\text{m}^{-1},\ \gamma_{AFM} = 0.08\times 10^{-3}\,\text{J m}^3/(\text{C}^2\,\text{A}^2). \tag{S.8b}$$

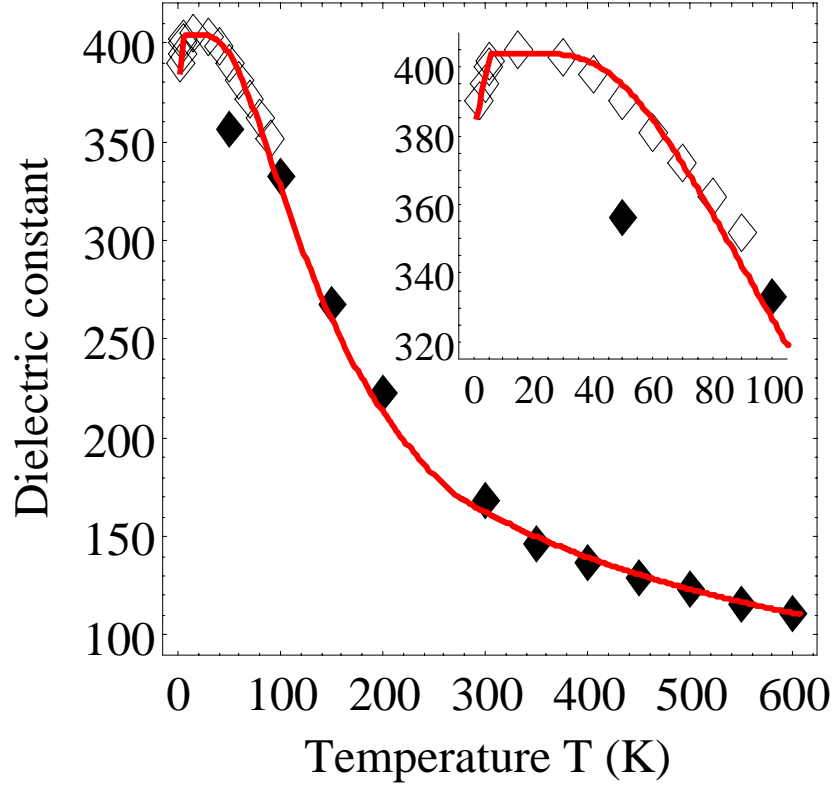

**Figure S5**. Comparison of fitting results with the experimental temperature dependence of dielectric constant in EuTiO$_3$. The open and filled diamonds represent experimental results collected from [3] and [4] respectively. The solid line is the fitting curve with phenomenological relation shown in Eq.(S.6). Inset shows a magnified view of the low-temperature region.

In order to estimate the remaining magnetic parameters, $\beta_M$ and $\lambda$, we have used the magnetic field dependences of magnetization $M$ and "antimagnetization" $L$. Using the expressions:

$$\alpha_M(T)M + \beta_M M^3 + \lambda L^2 M - \mu_0 H = 0,\qquad \alpha_L(T)L + \beta_L L^3 + \lambda L M^2 = 0. \tag{S.9}$$

and with the condition that at magnetic field about 1 Tesla, $M \approx L$ [3], we estimated $\beta_M = 0.8 \times 10^{-16}$ J m/A$^4$, $\lambda = 1.0 \times 10^{-16}$ J m/A$^4$.

In addition to the discussed fitting procedures, Fennie and Rabe [6] using *ab initio* calculations predicted that (001) EuTiO$_3$ thin films subjected to compressive epitaxial strain become ferromagnetic (FM) and ferroelectric (FE) simultaneously for strains exceeding 1.2%. Recently Lee *et al.* [7] demonstrated experimentally that an epitaxial strain indeed turns EuTiO$_3$ into multiferroic. In particular, epitaxially grown EuTiO$_3$ thin film on DyScO$_3$ substrate (corresponding to a tensile misfit strain of more that 1%) becomes FM at temperatures lower than 4.24 K and FE at temperatures lower than 250 K. Lee *et al.* explained the appearance of the ferromagnetism in EuTiO$_3$ thin film by the strong spin–lattice biquadratic magnetoelectric coupling.

The expansion coefficient $\beta_P$, representing spontaneous polarization of FE phase, could be estimated from the first principles calculations of Fennie and Rabe [6] and Lee *et al.* [7]. They reported the maximal spontaneous polarization of ~20–30 μC/cm$^2$ in compressed films and the critical misfit strains for the transition between AFM and FE+FM phases.

This fitting also allows us to estimate the electrostriction and magnetostriction constants, since the critical value of misfit strain governing the transition between paraelectric-AFM and FE-FM phases are calculated from the first principles by Lee *et al.* [7]. Since they found no intermediate phases (like ferroelectric-antiferromagnetic one), these findings allows us to determine unambiguously all the set of striction constants.